\documentclass[aip,cha,amsmath,amssymb,reprint]{revtex4-1}
\usepackage{color}
\usepackage{graphicx}
\usepackage{dcolumn}
\usepackage{bm}
\usepackage[mathlines]{lineno}
\usepackage{amsfonts}
\usepackage{amsmath}
\usepackage{sidecap}
\usepackage{float}
\usepackage{parskip}
\usepackage{grffile}

\usepackage{color}
\usepackage{hyperref}
\usepackage{hhline}
\usepackage{mathtools}
\usepackage{comment}
\usepackage[normalem]{ulem}
\makeatletter
\def\@eqnnum{{\normalsize \normalcolor (\theequation)}}
 \makeatother
\begin{document}

\title{Tiered synchronization in adaptive Kuramoto oscillators on simplicial complexes}
\author{Priyanka Rajwani}
\author{Ayushi Suman} 
\author{Sarika Jalan}
\email{sarika@iiti.ac.in} \affiliation{Complex Systems Lab, Department of Physics, Indian Institute of
Technology Indore, Khandwa Road, Simrol, Indore-453552, India}

\begin{abstract}
    An incorporation of higher-order interactions is known to lead an abrupt first-order transition to synchronization in otherwise smooth second-order one for pair-wise coupled systems. Here, we show that adaptation in higher-order coupling strength may yield completely different phenomena, notably, second-order transition to synchronization and tiered synchronization.Using the Ott-Antonsen approach, we perform rigorous theoretical calculations illustrating the origin of these emerging phenomena along with a complete description of all (un)stable states. Numerical simulations for limited-size networks are in agreement with the analytical predictions. These results would be important to comprehend dynamical behaviors of real-world complex systems with inherent higher-order interactions and adaptation through
    feedback coupling.
    \end{abstract}
    
\maketitle
\begin{quotation}
{\it{"It's not the strongest or most intelligent of the species that survives but the one that's most adaptable to change \cite{darwin1909origin}"  }} its essential to incorporate this paramount feature of adaptation while modeling dynamical behaviours of real-world systems. 
Adaptation in natural systems may arise due to environmental factors or self feedback to achieve a functional response or perform better.
Often signature of such adaptation is modeled as changes in the underlying network architecture or coupling strengths at which dynamical units of the system interact. Furthermore, recent drive to discern emerging behaviours of coupled dynamics on simplicial complexes has witnessed an unprecedented amount of work popularly establishing that an incorporation of higher-order interactions causes abrupt first-order transition to synchronization in a system which experiences smooth second-order synchronization in presence of only pair-wise coupling. This article merges these two domains of research; adaption and higher-order interactions, and demonstrates that an interplay of these can lead to diverse emerging phenomena such as tiered and second-order transitions to synchronization in addition to the first-order. 
 \end{quotation}

\paragraph{\bf{Introduction:}}
A system of weakly interacting units in a complex fashion often leads to synchronization \cite{osipov2007synchronization}.
Synchronization has been successfully observed and studied in various mathematical models and naturally occurring systems like flashing fireflies 
 \cite{buck1988synchronous}, crickets that chirp in unison 
  \cite{walker1969acoustic}, flocking of birds 
 \cite{attanasi2015emergence}, power-grids  \cite{article}, and Brain 
 \cite{adhikari2013localizing}.
The intriguing aspect here is that in absence of any interaction, system's units behave incoherently, and it is the existence or turning on of the interactions that lead to emerging phenomenon of coherence, often as a consequence of a phase transition. 
Such a phase transition from a disorder or incoherent state to synchronization or coherent state could be continuous \cite{Strogatz_Kura_rev}  or abrupt \cite{pazo2005thermodynamic}. The first-order abrupt jump from an incoherent to a coherent state, popularly referred as explosive synchronization (ES), has been reported to arise due to existence of degree-frequency correlation 
 \cite{gomez2011explosive}, 
 inertia term \cite{PhysRevLett.78.2104}, 
 bimodal intrinsic frequency distribution \cite{li2019clustering}, and inhibitory coupling \cite{jalan2019inhibition}. Recent addition to the systems properties which have exhibited to cause ES is existence of higher-order interactions. An incorporation of higher-order interactions in coupled Kuramoto oscillators model, which in presence of only pairwise coupling leads to smooth, second-order transition, has been shown to yield multistable states in the year 2011 \cite{tanaka2011multistable}. Since then, a series of studies including those by Skardal et al. \cite{Skardal_NatComm2020, Skardal_prl2019} and others \cite{SJ_Ayushi2022} have been conducted for systems consisting of higher-order interactions. 
 Based on these studies, the current state of knowledge is that an incorporation of higher-order interactions leads to explosive de(synchronization) in an otherwise smooth transition to synchronization in the same system having only pairwise interactions.

Furthermore, adaptation of coupling strength through order parameter allows us to model an interacting oscillators system having feedback coupling strength. Depending on the number of oscillators in the synchronized state, the coupling strength changes. For example, the emerging behavior of clapping in unison manifests that initially, the audience claps in an incoherent way, and due to the feedback sound of clapping the systems keep adapting, eventually yielding to clapping in synchronization. Such a form of the adaptation with global order parameter has been studied for pairwise interactions in Kuramoto model by Filatrella et. al. with an example of Josephson junction experiment \cite{filatrella2007generalized} and analytically explored by Zou and Wang \cite{zou2020dynamics}. Here, in this article, we study coupled Kuramoto oscillators with higher-order interactions on the impression of adaptation of the coupling strength. Although, an adaptation of order parameter in pair-wise interactions and triadic couplings, separately, are known to cause first-order phase transition. An interplay of these, as we discover here, can suppress explosive synchronization and indeed makes the system to follow the well-known continuous route to synchronization. In between the abrupt and continuous transition, an unconventional character of transition transpires, in which two tiered of the stable states i.e. bistability between a weakly synchronized state and strongly synchronized state exists after the curve folds over itself twice through a saddle-node bifurcation. Such a state is referred to as tiered synchronization and was earlier reported for systems
 having heterogeneous time-delayed couplings in simplicial complexes \cite{skardal2022tiered}. 
 A similar form referred to as Bellerophon was shown for pair-wise coupled Kuramoto oscillators with bimodal frequency distribution \cite{li2019synchronization}.

\paragraph{\bf{Model and Analytical derivation:}} We consider an extension of the Kuramoto model \cite{Strogatz_Kura_rev} by accounting for 1-simplex (pair-wise) and 2-simplex (triadic) interactions using sinusoidal couplings for $N$ number of oscillators with 1-simplex and 2-simplex coupling strength $K_1$ and $K_2$, respectively. Adaption has been put in the system by multiplying both the dyadic and triadic couplings with the global order parameter of the system. Two parameters $\alpha$ and $\beta$ control the extent of the adaptation, i.e., the degree of feedback through other oscillators. The individual phase of the oscillators then follow an $N$ coupled differential equation, 
\begin{align} \label{1}
\dot{\theta_i}=\omega_i&+\frac{{K_1}{r_1^\alpha}}{N}\sum_{j=1}^{N} 
         \sin(\theta_j-\theta_i)\nonumber  \\
        &+\frac{{K_2}{r_1^\beta}}{N^2}\sum_{j=1}^{N}  \sum_{k=1}^{N}\sin(2\theta_j-\theta_k-\theta_i)    
\end{align}
To characterize the macroscopic dynamics of this model, two complex-valued order parameters $z_1$ and $z_2$ are introduced
   \begin{equation} \label{order_par}
       z_p={r_pe^{i\psi_p}=\frac{1}{N}\sum_{j=1}^{N}e^{pi\theta_j}}
   \end{equation}
Here $z_1$ is the centroid of all the $N$ oscillator points ($e^{\iota\theta}$) in a complex plane and its real part $r_1 (0 \leq r_1 \leq 1)$ measures the extent of synchronization in the oscillator system with the quantity $\Psi_1$ being the mean phase of all the points. For a complete synchronized state $r_1=1$, and if all the oscillators are uniformly distributed over the unit circle, $r_1=0$. Though $z_1$ is sufficient to characterize a system of oscillators interacting in a pair-wise fashion,  due to the inclusion of higher harmonics in the coupling function, an order-2 parameter $z_2$ is also required. $z_2$ is the centroid of $N$ dynamic points $e^{2\iota\theta}$ in the complex plane. 
These definitions of order parameters allow us to write the model Eq.~\ref{1} such that the $N$ phase oscillators do not interact directly with each other but as if placed in a mean force field created by all the oscillators including the one in consideration. This mean field equation yields the stability condition for phase locking.

\begin{align}\label{mean_field}
\dot{\theta_i}=\omega_i&+{K_1}{r_1}^{\alpha+1}\sin(\psi_1-\theta_i)\nonumber \\
         &+{K_2}{r_2}{r_1}^{\beta+1}\sin(\psi_2-\psi_1-\theta_i) 
 \end{align}
 
If the oscillators are to synchronize, the steady state will be reached when the locked oscillators move with a certain common instantaneous frequency $\Omega$ which is the same as the mean phase of the locked oscillators. If we rotate the frame with this same angular frequency and choose the origin of the frame correctly, the mean-field will deliver the condition for phase locking such as $|{\frac{\omega-\Omega}{K_1r_1^{\alpha+1}+K_2r_2r_1^{\beta+1}}}|\leq 1$. Thus, a large spread in the intrinsic frequencies allows for a little fraction of the oscillators to form phase locking for a given coupling strength. The frequency distribution considered in this study is taken to be Lorentzian {\large$g(\omega)=\frac{\Delta}{\pi[(\omega-\omega_0)^2+\Delta^2]}$} (unless stated otherwise) with a peak at $\omega_0=0$ and has a standard deviation $\Delta=1$. 

\begin{figure*}[t!]
\begingroup
\begin{tabular}{c c c}
\includegraphics[width=0.33\textwidth]{images/fig4_1.eps}
\includegraphics[width=0.32\textwidth]{images/fig1.eps}
\includegraphics[width=0.321\textwidth]{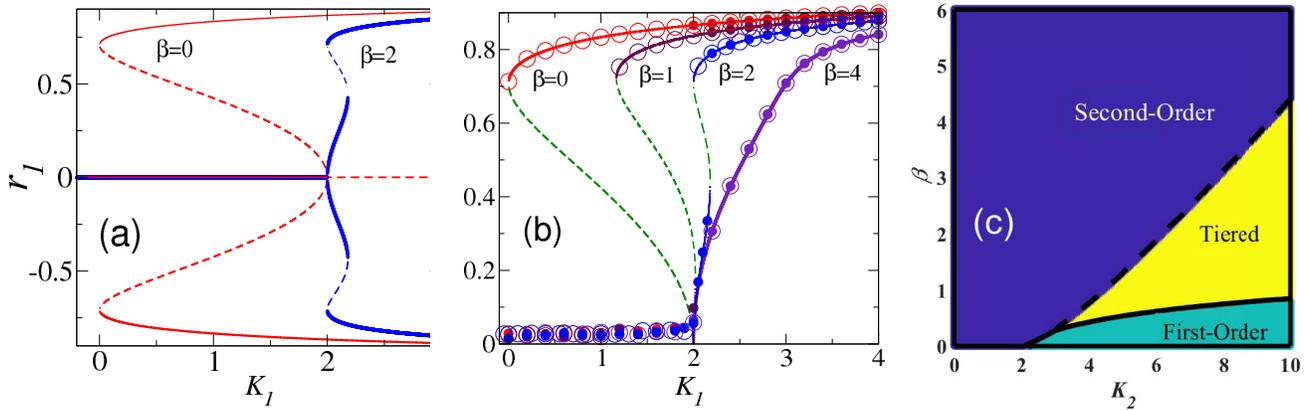}
\end{tabular}
\endgroup
\caption{
Order parameter $r_1$ as a function of $K_1$ for different values of $\beta$ and $\alpha=0$.
(a) is zoomed/{\bf elaborate} version of (b) and for selected values of $\beta$ and provides a clearer depiction of the transition of $r_1$ from abrupt to tiered to continuous synchronization. (b) The solid and open circles correspond to the numerical simulations of the system with adiabatically increasing $K_1$ in the forward and backward directions, respectively for $N=10^3$ oscillators. The solid and dashed lines in (a) and (b) correspond to the stable and unstable state, respectively as predicted analytically from Eq.~\ref{9}. $K_2=8$. (c) Parameter space ($K_2$ vs. $\beta$) predicted analytically by calculating roots of Eq.~\ref{K1}. For different $K_2$ and $\beta$ values, regions of first-order (abrupt), tiered, and second-order (continuous) path of synchronization are depicted.
} 
\label{Fig_1}
\end{figure*}
For dimensionality reduction of the $N$ oscillator system, Eq.~\ref{mean_field} can be composed in terms of the complex order parameters such as
\begin{align} \label{4}
    \dot{\theta_i}= \omega_i&+\frac{K_1r_1^\alpha}{2\iota}{(z_1e^{-i\theta_i}-z_1^*e^{i\theta_i})} \nonumber \\
    &+\frac{K_2r_1^\beta}{2\iota}{(z_2z_1^*e^{-i\theta_i}-z_2^*z_1e^{i\theta_i})}     
\end{align}
In the continuum limit $N\rightarrow\infty$, the state of the system can be described by a density function $f(\theta,\omega,t)$, which specifies the density of the oscillators with phase lying between $\theta$ and $\theta+\delta\theta$ having a natural frequency between $\omega$ and $\omega+\delta\omega$ at time $t$. Whereas the number of oscillators in the system is conserved, the density function satisfies the continuity equation, 
   \begin{equation} \label{continuity}
    \frac{\delta{f}}{\delta{t}}+\frac{\delta({f\dot{\theta}})}{\delta\theta}=0
    \end{equation}
Considering that the oscillators natural frequency is drawn from a distribution $g(\omega)$, the density function can be expanded into the Fourier series as 
  \begin{equation} \label{fourier}
     f(\theta,\omega,t)=\frac{g(\omega)}{2\pi}\left[\sum_{n=-\infty}^{\infty}f_n(\omega,t)e^{in\theta}\right]
  \end{equation}
with $f_{-n}=f_n^*e^{-in\theta}$, $f_n(\omega,t)$ being the $n^{th}$ Fourier coefficient.
Further, practicing the Ott-Antonsen \cite{Ott_Antenson2008} assumption that all the Fourier modes decay geometrically, i.e. $f_n(\omega,t)=\upsilon^n(\omega,t)$, plugging this and Eq.~\ref{4} into the continuity Eq.~\ref{continuity}, the dynamics of the network collapse into a complex one-dimensional manifold.
   \begin{align} \label{ott_ant}
   \dot{\upsilon}&+i\upsilon\omega-\frac{K_1{r_1^\alpha}z_1^*}
   {2}+\frac{K_1{r_1^\alpha}z_1{\upsilon^2}}{2}\nonumber \\
   &+\frac{K_2{r_1^\beta}z_2z_1^*\upsilon^2}{2}-\frac{K_2r_1^\beta{z_2^*z_1}}{2}=0 
   \end{align}
In the continuum limit $N\rightarrow\infty$, we have $z_p=\int_{-\infty}^{\infty}\int_{0}^{2\pi}{f(\theta,\omega,t)e^{ip\theta}d\theta{d\omega}}$, which after inserting the Fourier series expansion of $f(\theta,\omega,t)$this reduces to $z_1=\int_{-\infty}^{\infty}g(\omega){\upsilon^*}d\omega$ and $z_2=\int_{-\infty}^{\infty}g(\omega){\upsilon^*}^{2}d\omega$.If we consider the frequency distribution $g(\omega)$ to be Lorentzian {\large$g(\omega)=\frac{\Delta}{\pi[(\omega-\omega_0)^2+\Delta^2]}$}with mean $\omega_0=0$ and spread $\Delta=1$. The integral of ${z_1}$ and  ${z_2}$ can be calculated using Cauchy's Residue Theorem by contour integration in a negative half-plane, yielding $z_1={\upsilon^*}(\omega_0-i\Delta,t)$ and $z_2=({\upsilon^*}(\omega_0-i\Delta,t))^{2}$. After, inserting these real part of Eq.~\ref{ott_ant}  in terms of $r_1$ can be reduces to,
   \begin{equation}\nonumber
    \dot{r_1}=-r_1+\frac{K_1{r_1^\alpha}(r_1-r_1^3)}{2}+\frac{K_2{r_1^\beta}(r_1^3-r_1^5)}{2}
   \end{equation}
   In the steady state, $\dot{r_1}=0$ implies that,
   \begin{equation}\label{9}
    0=-r_1+\frac{K_1{r_1^\alpha}(r_1-r_1^3)}{2}+\frac{K_2{r_1^\beta}(r_1^3-r_1^5)}{2}
   \end{equation}
After performing a stability analysis for Eq.~\ref{9} we learn that in the case of no adaptation in pair-wise interactions i.e. $(\alpha=0)$,$r_1=0$ solution is stable and unstable for $K_1<2$ and  $K_1>2$, respectively.

\paragraph{\bf {Tiered to second-order synchronization:}}
 First, we study the adaptation employed only in 2-simplex coupling i.e. $\beta \ne 0$ and $\alpha=0$. For this part, the dimension reduced Eq.~(\ref{9}), considering any non-zero value for the spread $\Delta$, yields,
   \begin{equation}\nonumber
    0=-r_1\Delta+\frac{K_1}{2}(r_1-r_1^3)+\frac{K_2{r_1^\beta}}{2}(r_1^3-r_1^5)
   \end{equation}
which can be solved exactly to find the critical points for the onset of the synchronization. By making the first derivative of $K_1$ with respect to $r_1$ equal to zero, one  achieves 
\begin{equation}\label{K1}
    K_{1f}=\frac{2(1+r_1^2)}{(1-r_1^2)^2}\Delta-K_2(\beta+3)r^{\beta+2}
\end{equation}

Since synchronization occurs at $r_1=0$, the critical coupling turns out to be independent of $K_2$ and $\beta$ and only depends on $2\Delta$ ($\Delta=1$ for all the results presented here). Therefore, the critical coupling strength at which forward transition remains the same for any value of $K_2$, however,  the nature of transition changes depending upon the sign of the double derivative of $K_1$ with respect to $r_1$. 
   \begin{equation}\nonumber
    \frac{d^2K_1}{dr_1^2}=12-K_2(\beta+3)(\beta+2)r_1^{\beta}
   \end{equation}
Here, for any positive finite value of $\beta$, $\frac{d^2K_1}{dr_1^2}$ is positive signifying a local minimum that corresponds to a supercritical pitchfork bifurcation in $r_1-K_1$ space. Here, $r_1=0$ is a stable state, which changes into an unstable state at the bifurcation point ($K_{1f}=2$). 
For the nonzero exponent $\beta$ as $K_1$ increases a bit further, the system encounters a pair of saddle-node bifurcation arising due change of the solution of Eq.~\ref{K1}. As a consequence of the interplay of the exponent $\beta$ and $K_2$, the system leaves the supercritical synchronization path (weakly synchronized state) and jumps to a strongly synchronized state. This route to synchronization which commences as second-order followed by a discontinuous jump is referred to as tiered synchronization \cite{skardal2022tiered}. The stability or bifurcation diagram is depicted in Fig.~\ref{Fig_1}(a) with the solid and dashed lines depicting stable and unstable states, respectively. For the initial condition corresponding to the $r_1=0$ state, i.e. starting with a state in which oscillators are homogeneously distributed in phase, and  $\beta=0$ (red curve), the transition is strictly of first-order (sub-critical pitchfork bifurcation), i.e., the system abruptly goes from the incoherent to the coherent state and always forms a hysteresis when traveled adiabatically in the backward direction on the $K_1$ axis. On the other hand, for the non-zero value of the exponent $\beta$, the trajectory does not directly jump to $r_1=1$ state but transits first to a finite positive value of $r_1$ followed by an abrupt jump (tiered synchronization). In Fig.~\ref{Fig_1}(a) for $\beta=2$ the solid line (blue curve) illustrates the occurrence of tiered synchronization; initially, the system experiences super-critical pitchfork bifurcation and then jumps to the maximum possible phase coherence state. 

\begin{figure*}
\includegraphics[width=0.9\textwidth]{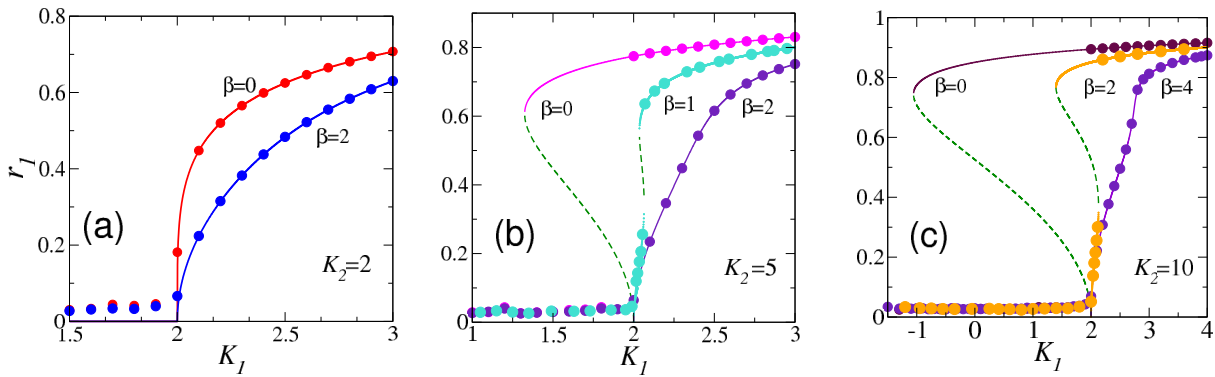}
    \caption{Synchronization transition of $r_1$ for different values of 2-simplex coupling strength $K_2$.  After the critical value of $K_2$, as exponent $\beta$ increases the transition to synchronization shifts from first-order to second-order via tiered synchronization. Other parameters are the same as in the previous figure. Solid circles correspond to the numerical simulation of the
system with adiabatically increasing 1-simplex coupling $K_1$ in the forward directions for $N = 10^3$ oscillators with random initial conditions. }
    \label{fig2}
\end{figure*}

\paragraph{{\bf Numerical calculations:}} Numerical simulations are  performed for the Eq.~\ref{mean_field} which is mean field form of the model Eq.~\ref{1} by changing the pair-wise interaction coupling $K_1$ in the forward and backward direction. For each $K_1$ value, the simulation runs for $3\times10^4$ iteration with the time step $dt=0.01$. After removing an initial transient. $r_1$ is averaged over $10^4$ iterations for $N=10^3$ oscillators.  
The numerical results are plotted along with the analytically solved Ott-Antonsen Eq.~\ref{9}.
For the case of adaptation in 2-simplex interactions, i.e., exponent ($\beta\ne0$)  with no adaptation in the pair-wise interactions i.e. ($\alpha=0$), the system endeavors to adapt the value of 2-simplex coupling strength $K_2$ as slowly as the value of $\beta$ increases. The factor $r_1^\beta$ changes the solutions of the model Eq.~\ref{1} and dynamics of the system. For the value of $\beta=0$ system manifests the pitchfork bifurcation at $K_{1f}=2$ (supercritical and sub-critical for $K_2<2$ and $K_2>3$, respectively) \cite{Skardal_NatComm2020}. However, for non-zero values of $\beta$ and $K_{1f}>2$ system leaves off $r_1=0$ state and struggles to achieve synchronized state. The type of synchronization rests upon pair-wise or higher-order interaction strengths. For any value of $0<r_1<1$ (Eq.~\ref{order_par}), $K_2$ multiplies with $r_1^\beta$, thereby leads to a decrease in effective coupling strength $K_2$. Therefore, the pair-wise interaction term dominates the system. As $K_1$ increases, the system gradually favors the phase synchronization, and the value of $r_1$ increases, thereby increasing the effective 2-simplex coupling strength $K_2$ and making higher order coupling strength term take command.  As the higher-order couplings favor abrupt (first-Order) synchronization, the system supports an abrupt maximum phase coherence state depending upon the value of $\beta$ along with the value of $K_2$ as the adaptation factor $r_1^\beta$ changes the number of solutions of the model Eq.~\ref{1}. Ergo, we obtain the tiered as well as continuous (second-order) synchronization. 

For $K_2=8$, an increase in $\beta$ further reduces the effective coupling strength of 2-simplex interactions, and the system follows a continuous synchronization path (Fig.~\ref{Fig_1}(b)). For the value of $\beta=2$, initially, the pair-wise interactions term controls the dynamical behaviour of the system, yielding the continuous second-order transition to synchronization. Thereupon, with an increase in $r_1$, as the effective 2-simplex coupling strength increases, it makes the system to experience an abrupt jump into a maximum phase coherence state by creating two saddle nodes and leaving the supercritical bifurcation; The entire process known as tiered synchronization, i.e., a bistability state exists between a weakly synchronized state (small value of $r_1$) and strongly synchronized state after the curve folds over itself twice through a saddle nodes bifurcation \cite{skardal2022tiered}. For the value of $\beta \geq 3$, the contribution from the 2-simplex coupling strength term reduces more and the pair-wise interactions term dominates. Consequently, the system follows the continuous route to synchronization.  When the 2-simplex coupling strength term dominates, the system has already attained the maximum phase coherence state through the continuous path. In Fig.~\ref{Fig_1}(c) the parameter space ($K_2$ vs. $\beta$) is predicted analytically and plotted by calculating the number of roots of Eq.~\ref{K1}. In the first-order phase transition, the tangent of curve $dK_1/dr_1$ is equal to zero at two points  ($r_1 \ge 0$) (Fig.~\ref{Fig_1}(a)). Therefore, Eq.~\ref{K1} admits two roots for a particular combination of $K_2$ and $\beta$ yielding first-order (abrupt) synchronization (light grey, sea green) region. Similarly, for three roots being admissible one gets tiered synchronization (dark grey, yellow) region, and for the situation when Eq.~\ref{K1} has only one root at $r_1=0$. second-order(continuous) synchronization (black, blue) region is obtained. This clearly highlights the role of an interplay of $K_2$ and exponent $\beta$ in admitting different dynamical behaviours, particularly three distinct regions.

\begin{figure}
\includegraphics[width=0.35\textwidth]{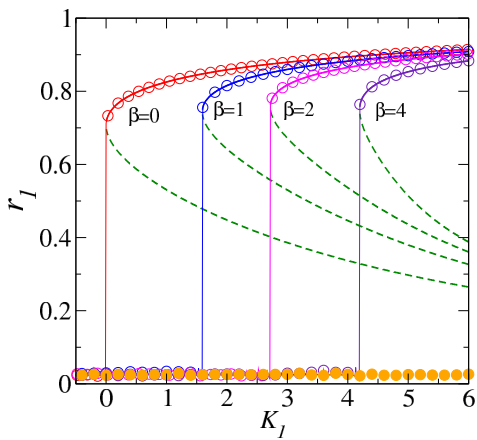}
    \caption{Presence of adaptation in 1 and 2-simplex interactions. Order parameter $r_1$ is the plotted as a function of $K_1$ for different $\beta$ values. As $\beta$ increases, the backward transition point shifts toward the higher positive value of $K_1$. Open and closed circles correspond to numerical simulation for adiabatically decreasing and increasing, respectively, $K_1$ for $N=10^4$ oscillators. Other parameters are $K_2=8$, and $\alpha=1$}
    \label{fig3}
\end{figure}

Fig.~\ref{fig2} further elaborates the effect of change in the value of $K_2$. For a threshold value of $K_2(<3$ (Fig.~\ref{fig2}(a)) below which no tiered synchronization is obtained, the system follows only the continuous path to synchronization for all the values of $\beta$. Furthermore, as the 2-simplex coupling strength $K_2$ increases, a larger value of $\beta$ is required for the system to manifest a second-order phase transition. Also $K_2$ plays a significant role in enhancing the range of $\beta$ for which tiered synchronization is obtained (Fig.~\ref{Fig_1}(c)). For instance, as illustrated by Fig.~\ref{fig2}(b), for $K_2=5$ system attains the second-order phase transition for $\beta=2$, whereas for $K_2=10$ system attains the second-order phase transition for $\beta=4$ and follows a tiered synchronization for lower values of the $\beta$ (Fig.~\ref{fig2}(c)) It follows that for higher $K_2$ values, for larger $\beta$ values only the effective strength of the 2-simplex interactions ($r_1^\beta K_2$) is small, and therefore the pair-wise interaction term takes the command yielding second-order transition to synchronization. Ergo, abrupt to the tiered to the continuous route of synchronization is determined together by $K_2$ and  $\beta$  due to the effective adaptation parameter $r_1^\beta$ that changes the dynamics of the system.

\paragraph{\bf Impact of adaptation in pair-wise interactions:}
Fig.~\ref{fig3} demonstrates the effect of the introduction of adaptation in the pair-wise interactions i.e. $\alpha\neq 0$. The studies on pair-wise interactions with adaptation through global order parameter by Filatrella et al. \cite{filatrella2007generalized} show the hysteresis behaviour for $N=1000$ oscillators and the same model further carried out by Zou and Wang \cite{zou2020dynamics} performed the stability analysis and show the detailed bifurcation diagram. In consequence, for $\alpha>1$ only abrupt desynchronization is observed because the adaptation parameter $r_1^\alpha$ in the pair-wise interactions makes $r_1=0$ solution stable for all the values of $K_1>0$ and the unstable state moves in the upward direction (for small value of $\alpha$), also asymptotically approaches to incoherent state for large value of $K_1$. Simultaneously, as the higher-order coupling term comes along with adaptation by global order parameter($r_1^\beta$), this shows that for a fixed value of the 2-simplex coupling strength, $K_2$ as the exponent $\beta$ increases the value of the backward transition coupling point shifts towards the higher positive 1-simplex coupling strength $K_1$. The results claimed here are true for thermodynamic limit $N\rightarrow\infty$, In the case of a finite number of oscillators when the asymptotically approaching unstable state and stable state for $r_1=0$ solution meet up there is a forward jump in the numerical results that are not shown by analytical produce bifurcation diagram for $N\rightarrow\infty$.

\paragraph{\bf{Conclusion and future prospects:}} We have discovered that Adaptation of global order parameter in simplicial complexes can lead to tiered synchronization. We have presented rigorous analytical calculations for accessing the changes in the order parameters with respect to interaction strengths.
 It was delineated that tiered synchronization can be observed for time-delayed interaction only \cite{skardal2022tiered}. This article demonstrates that adaptation through the global order parameter in 2-simplex interactions can also yield tiered synchronization.
We have demonstrated that an interplay between the 2-simplex adaptation exponent ($\beta$) and 2-simplex coupling constant ($K_2$) is of prime importance since the former controlling the latter gives birth to the tiered synchronization. 
Furthermore, upon switching on the adaptation in the pair-wise couplings as well, synchronization gets killed in the forward direction which is in complete contrast to the zero adaptation case, and in fact, resembles the case of an absence of the pair-wise coupling \cite{Skardal_prl2019}. Upon setting the initial conditions to a complete or partially synchronized state, the coupled dynamics witnesses an abrupt desynchronization transition to an incoherent state. This article only presents an analysis for globally connected system, whereas real-world complex systems have underlying network architecture. This study can be further extended to analyze the role of adaptation in higher-order interactions for various network models representing connections architecture of real-world systems.
We expect that the results presented here could be valuable in understanding and predicting dynamical behaviours of real-world complex systems such as the brain and other systems where adaptation \cite{van1997hebbian} or feedback coupling
 \cite{filatrella2007generalized} play crucial role in their functioning. Also, this study would help in understanding the origin of phase transition behaviours in systems where adaptation and higher-order interactions are at play.

\section{acknowledgement}
SJ gratefully acknowledges SERB Power grant SPF/2021/000136. 
We are thankful for the computational facility received from Department of Science and Technology (DST), Government of India, under FIST scheme (Grant No. SR/FST/PSI-225/2016).

\medskip
\bibliography{ref}
\end{document}